\documentclass[fleqn,12pt,twoside]{article}
\usepackage[headings]{espcrc1}

\readRCS
$Id: espcrc1.tex,v 1.2 2004/02/24 11:22:11 spepping Exp $
\usepackage{graphicx,fixmath}
\usepackage[figuresright]{rotating}

\newcommand{\AmS}{{\protect\the\textfont2
  A\kern-.1667em\lower.5ex\hbox{M}\kern-.125emS}}

\title{Azimuthal Anisotropy of Identified Hadrons in 200\,GeV\,Au+Au Collisions}

\author{M.\,Oldenburg\address[LBL]{Lawrence Berkeley National Laboratory,\\ 
        Nuclear Science Division, 
        One Cyclotron Road, Berkeley, CA 94720, USA}%
\ (for the STAR\footnote{For the full list of STAR authors and acknowledgments, see appendix `Collaborations' of this volume.} Collaboration)       
        }
\runauthor{M.\,Oldenburg% (for the STAR Collaboration)
}
\begin{document}

% typeset front matter
\maketitle

\begin{abstract} 

The azimuthal anisotropy parameter $v_2$ has been measured with high
statistics for identified hadrons in
$\sqrt{s_{NN}}=200\,\mathrm{GeV}$\,Au+Au collisions with the STAR
experiment. At high transverse momentum $(p_T)$ a strong $v_2$ for
$\pi^+ + \pi^-$ and $p +\overline{p}$ is observed. In the intermediate
$p_T$ region, number-of-constituent-quark scaling was tested to high
precision. A detailed comparison of $v_2$ for the multi-strange
hadrons $\phi$, $\Xi^- + \overline{\Xi}^+$, and $\Omega^- +
\overline{\Omega}^+$ with other particle species substantiates the
development of collectivity among partons in the early phase of the
collisions at RHIC.

\end{abstract}

\section{Introduction}

The study of ultra-relativistic heavy-ion collisions provides insight
into properties of very high density nuclear matter. Since the
observed particle distributions to first order only reflect the
conditions in the final state of the system, signatures originating
from the early stage are needed to conclude whether the system passes
through a partonic phase. This is one necessary step to identify the
predicted state of matter called the quark-gluon plasma; the other one
being the proof that the system is thermalized.

One observable which is sensitive to the early stage of the collision
is the second harmonic coefficient, $v_2$, of the Fourier expansion of
the azimuthal momentum distribution, called elliptic flow
\cite{flow}. In non-central collisions the initial spatial anisotropy
is transformed into an anisotropy in momentum-space if sufficient
interactions occur among the constituents within the system. Once the
system has expanded enough to quench the spatial anisotropy, further
development of momentum anisotropy ceases.

In the following we will discuss results on $v_2$ for identified
hadrons measured in $\sqrt{s_{NN}}=200\,\mathrm{GeV}$ minimum bias
Au+Au collisions from the STAR experiment at RHIC.

\section{$\mathbold{v_2}$ of identified hadrons at high $\mathbold{p_T}$}

\begin{figure}[htb]
%	\vspace*{-0.3cm}
	\begin{minipage}[h]{0.47\linewidth}
	\vspace*{-0.5cm}
		\raisebox{1.2cm}{
		\includegraphics[width=\linewidth]{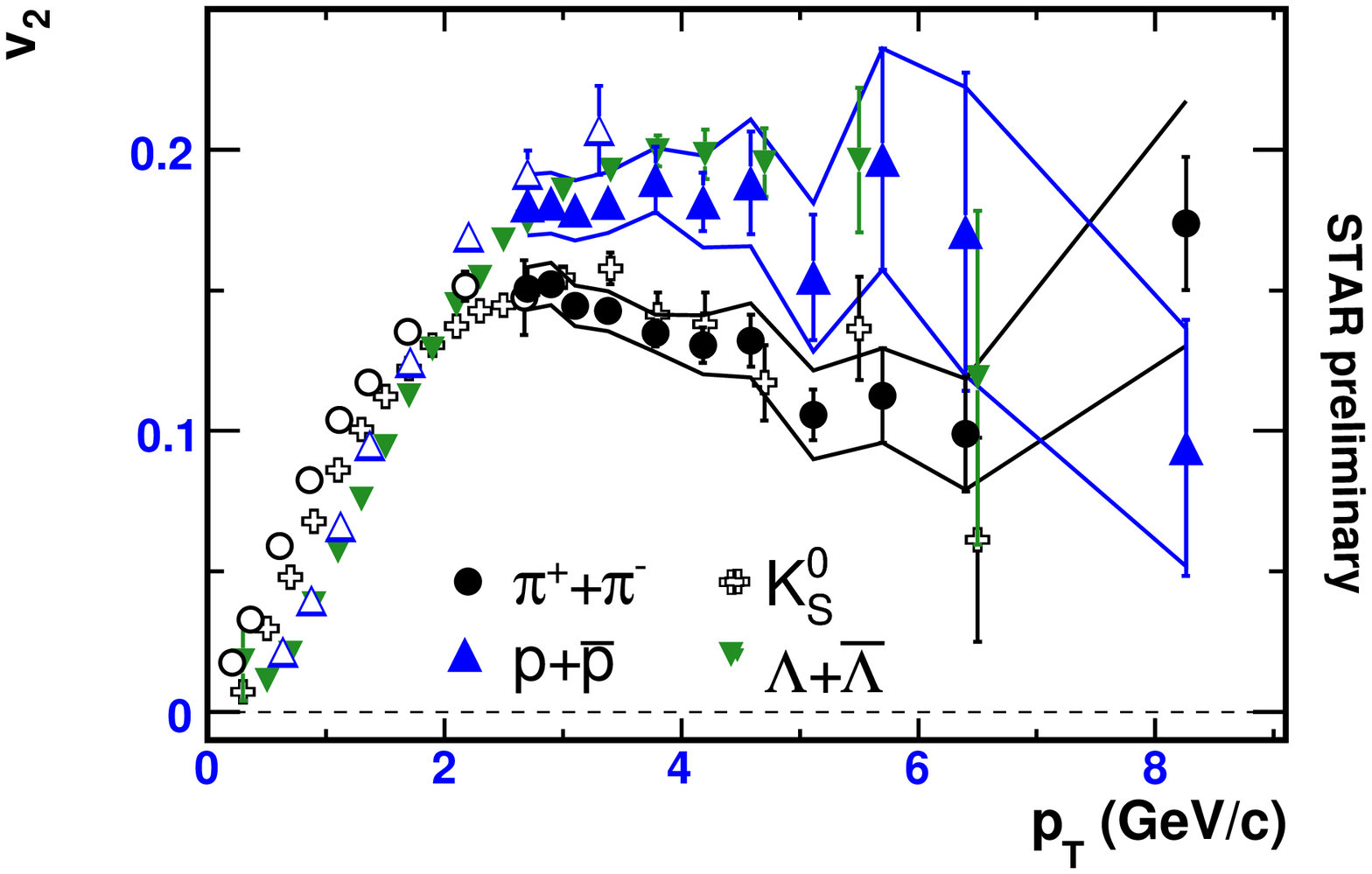}
}
\vspace*{-2.55cm}
		\caption{(left) Azimuthal anisotropy $v_2$ for $\pi^+
		+ \pi^-$ and $p +\overline{p}$ in 200~GeV minimum
		bias Au+Au collisions. $K^0_S$ and $\Lambda +
		\overline{\Lambda}\ v_2$ are
		shown for comparison.\label{fig:1}}
		\end{minipage}\hfill
	\begin{minipage}[h]{0.49\linewidth}	
	\includegraphics[width=\linewidth]{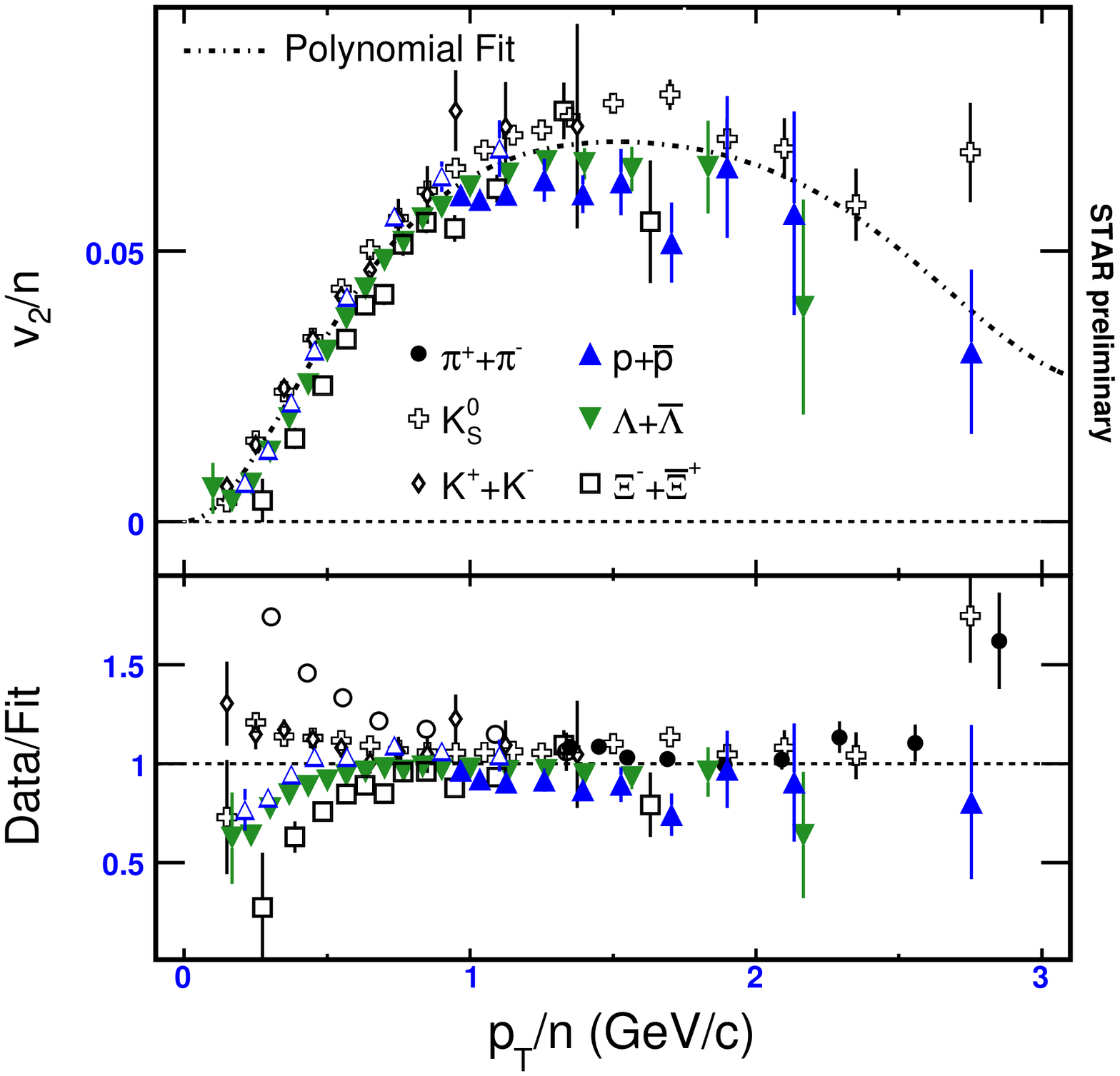}
	\end{minipage}
\vspace*{-0.7cm}
	\caption{(right) Measurements of scaled $v_2(p_T/n)/n$ for
		identified hadrons (upper panel) and ratio (lower
		panel) between the measurements
		and a polynomial fit through all data
		points except pions for 200~GeV minimum bias Au+Au
		collisions.\newline
		For both figures the open symbols
		for $\pi^+ + \pi^-$ and $p +\overline{p}$ were taken
		from \cite{PHENIX_data}.\label{fig:2}} 
		\end{figure}

With the high statistics available from RHIC's run IV the coverage for
identified hadron $v_2$ was extended up to $p_T = 9\,\mathrm{GeV}/c$
by disentangling the different contributions to the overall
$v_2^\mathrm{tot}(\mathrm{d}E/\mathrm{d}x)$, as shown in
Fig.\,\ref{fig:1}.  To suppress non-flow effects, particles were
correlated to the reaction plane determined in the opposite
$\eta$-subevent. The difference of this $v_2$-measurement from the
results of the method where the event plane is measured over the full
acceptance, is an estimate of one of the contributions to the
systematic error; another factor in the systematic error is based on
the estimated level of contamination of protons by kaons. The
systematic errors are shown as bands in Fig.\,\ref{fig:1}, while the
error bars include statistical uncertainties only. At low $p_T$ the
systematic errors were estimated to be less than 10\,\%
\cite{longFlowPaper}.

While the proton $v_2$ saturates at $p_T \sim 3\,\mathrm{GeV}/c$ the
measurements for $\pi^+ + \pi^-$ show a decline starting from their
maximum at about $p_T = 3\,\mathrm{GeV}/c$. The flow of these
non-strange hadrons is strikingly similar to that of the strange
particles $K^0_S$ and $\Lambda + \overline{\Lambda}$. The mesons and
baryons fall into separate groups and this behavior seems to extend
out to rather high transverse momentum. Most surprisingly, even at the
highest $p_T$ measured a large $v_2$ is still observed. This appears
to be in contradiction to parton energy loss models \cite{Xin-Nian}
and multi-component calculations \cite{Denes}, which both predict a
smaller $v_2$ at and above $p_T \sim 6\,\mathrm{GeV}/c$. At these high
momenta the remaining $v_2$ value should be essentially driven by the
path-length dependence of jet quenching only \cite{consti}.

\section{Number-of-constituent-quark scaling of $\mathbold{v_2}$\label{ncqOldi}}

To further study the observed grouping into mesons vs.\ baryons, the
top of Fig.\,\ref{fig:2} shows the scaled $v_2/n$ for identified
hadrons over a broad range of scaled transverse momentum $p_T/n$. Here
$n$ is the number of constituent quarks (NCQ) of a given
hadron. Figure~\ref{fig:2}, bottom displays the ratio between the
measurements and a polynomial fit to all data. At low $p_T/n$ $(<
0.75\,\mathrm{GeV}/c)$ the observed deviations from the fit follow a
mass-ordering which is expected from hydrodynamic flow. At higher
$p_T$, all $v_2/n$ measurements are very close to the common `mean
value'. While the shown errors include statistical uncertainties only,
it is still striking to see that the similarity of the scaled $v_2$
extends out to $p_T/n$ as high as $2\,\mathrm{GeV}/c$. It seems that
all mesons fall above the fit while all baryons fall below.

The observation that the $v_2(p_T/n)/n$ is similar for all the
different particle species, strongly supports quark coalescence (see
for example \cite{NCQMolVol}) as the dominant process of hadronization
in the intermediate $p_T$ region. With the available statistics small
deviations from exact NCQ-scaling are now detectable, which were
expected even for simple recombination models \cite{NCQMolVol}.

\section{$\mathbold{v_2}$ of multi-strange hadrons}

Compared to $\pi^+ + \pi^-$, $K^+ + K^-$, and $p + \overline{p}$
\cite{piKp}, measurements of particle spectra at RHIC have shown that
the multi-strange hadrons $\phi$ and $\Omega$ are less affected by the
hadronic phase \cite{strangeProd} than by the partonic phase.
%While hadrons like $\pi^+ + \pi^-$, $K^+ + K^-$, and $p + 
%\overline{p}$ show a similar behavior of high mean collective velocity
%$\langle\beta_T\rangle$ and moderate kinetic freeze-out temperature
%$T_{\mathrm{fo}}$ (going from $\langle\beta_T\rangle = 0.59\,c$ for
%central to $0.36\,c$ for peripheral collisions and a change from
%$T_{\mathrm{fo}} = 89\,\mathrm{MeV}$ to $129\,\mathrm{MeV}$
%\cite{piKp}) for multi-strange hadrons $\langle\beta_T\rangle=0.42\,c$ 
%and $T_{\mathrm{fo}}=182\,\mathrm{MeV}$ \cite{strangeProd}. Due to the
%measured high freeze-out temperature the interaction of multi-strange
%hadrons with non-strange particles cannot be as strong as interactions 
%among non-strange particles themselves. 
It has been demonstrated \cite{omegaFlow} that the $\Xi^- +
\overline{\Xi}^+$ and $\Omega^- + \overline{\Omega}^+$ baryons possess
a significant amount of $v_2$, which suggests that this azimuthal
anisotropy is generated at an early and partonic stage of the system's
evolution.

For the $\phi$ meson, earlier measurements disfavored kaon coalescence
as the dominant channel for $\phi$ production at RHIC energies
\cite{phi}. Also its lifetime is long compared to the lifetime of the
fireball. Hence the $\phi$ is a useful probe of the early stage of the
system as well. The $v_2$ of the $\phi$ is particularly interesting,
since the $\phi$-mass of $1019.5\,\mathrm{MeV}/c^2$ is similar to the
mass of the proton $(938.3\,\mathrm{MeV}/c^2)$.

\begin{figure}[htb]
\vspace*{-1cm}
	\begin{minipage}[c]{0.28\linewidth} 
		\caption{Azimuthal
		anisotropy $v_2$ for strange hadrons (left)
		and multi-strange hadrons (right) in 200~GeV minimum
		bias Au+Au collisions. The dashed lines show a common fit
		\cite{Xin} to the $K^0_S$ and $\Lambda +
		\overline{\Lambda}$ data. Hydrodynamic model
		calculations
		\cite{hydro} are shown as shaded areas.
		\label{fig:3}}
			\end{minipage}\hfill
	\begin{minipage}[c]{0.71\linewidth}\hspace*{0.5cm}
\includegraphics[width=\linewidth]{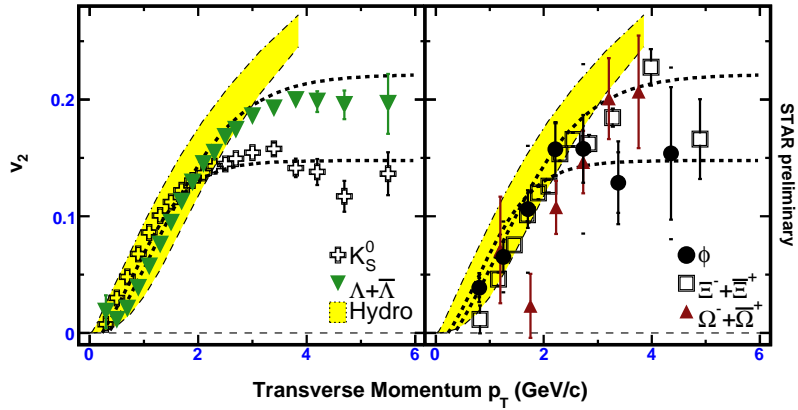}		
	\end{minipage}	
\end{figure}

The new high statistics measurements of azimuthal anisotropy $v_2$ of
multi-strange hadrons show strong elliptic flow for all studied
particles: $\Xi^- + \overline{\Xi}^+$, $\Omega^- +
\overline{\Omega}^+$, and $\phi$ (see Fig.\,\ref{fig:3}, right). The
error bars contain statistical uncertainties only. For the
$\phi$-measurement systematic errors obtained by the comparison of two
different methods are also shown. A common fit to the results obtained
for $K^0_S$ and $\Lambda + \overline{\Lambda}$, as depicted in
Fig.\,\ref{fig:3}, left and motivated by number-of-constituent-quark
scaling \cite{Xin}, was used to compare the $v_2$-data of
multi-strange hadrons to those of the other particles. A detailed
comparison, carried out by calculating $\chi^2/\mathrm{n.\,d.\,f.}$
for the difference between the measurements and the fit function (see
Tab.\,\ref{table:1}), suggests that $\Xi^- + \overline{\Xi}^+$ and
$\Omega^- + \overline{\Omega}^+$ indeed flow as strongly as the other
baryons, while the $\phi$ favors the similarity to other mesons. Note
that the fit function does not reproduce the data perfectly due to
small deviations from the exact number-of-constituent-quark scaling,
as discussed in Sec.\,\ref{ncqOldi}.

\vspace*{-0.15cm}
\begin{table}[htb]
	\begin{minipage}[h]{0.40\linewidth} \caption{Comparison of
	$v_2(p_T)$ measurements for multi-strange hadrons to a common
	fit \cite{Xin} to $v_2(p_T)$ of $K^0_S$ and $\Lambda +
	\overline{\Lambda}$.  \label{table:1}}
	\end{minipage}\hfill
	\begin{minipage}[c]{0.56\linewidth}

\newcommand{\m}{\hphantom{$-$}}
\newcommand{\cc}[1]{\multicolumn{1}{c}{#1}}
\renewcommand{\arraystretch}{1.2} % enlarge line spacing
\begin{tabular}{@{}ccc}
\hline
$p_T/n$&$\chi^2/\mathrm{n.\,d.\,f.}$& $\chi^2/\mathrm{n.\,d.\,f.}$ \\
 $> 0.75\,\mathrm{GeV}/c$  &for NCQ\,=\,2&for NCQ\,=\,3 \\
\hline
$\Xi^-+\overline{\Xi}^+$&82.7/6&{\bf 27.4/6} \\
$\Omega^-+\overline{\Omega}^+$&4.1/3&{\bf 2.1/3} \\
$\phi$&{\bf 2.0/5}&7.4/5 \\
\hline
\end{tabular}\\[2pt]
	\end{minipage}	
\end{table}
\vspace*{-0.35cm}

Since the $\phi$ flows like a meson and not as strongly as baryons,
e.\,g.\ the proton with its similar mass, it is found that the
observed scaling is a meson-baryon effect and not a mass effect. The
observed strong flow of multi-strange hadrons substantiates the
development of collectivity among partons at RHIC.

\section{Summary} 

We have presented the azimuthal anisotropy parameter $v_2$ for
identified particles in 200\,GeV minimum bias Au+Au
collisions. Elliptic flow of $p+\overline{p}$ and $\pi^+ + \pi^-$ was
measured up to $p_T=9\,\mathrm{GeV}/c$, where these particles still
show a strong $v_2$ signal. The observed meson-baryon grouping at
intermediate $p_T$ suggests parton coalescence to be the dominant
process of hadronization in that region, even though small deviations
from ideal number-of-constituent-quark scaling are observed. This
scaling is a meson-baryon effect and not a mass effect. Finally, the
multi-strange hadrons $\phi$, $\Xi^- + \overline{\Xi}^+$, and
$\Omega^-+\overline{\Omega}^+$ flow as strongly as the other mesons
and baryons, which confirms partonic collectivity at RHIC. The
remaining item to address for the discovery of a quark-gluon plasma is
the thermalization of the system.


\begin{thebibliography}{99}
\bibitem{flow} A.\,M.~Poskanzer and S.\,A.~Voloshin, Phys.\,Rev.~{\bf C\,58} (1998) 1671.
\bibitem{PHENIX_data} S.~Adler \emph{et\,al.} (PHENIX Collaboration), Phys.\,Rev.\,Lett.~{\bf 91} (2003) 182301.
\bibitem{longFlowPaper} J.~Adams \emph{et\,al.} (STAR Collaboration), Phys.\,Rev.~{\bf C\,72} (2005) 014904.
\bibitem{Xin-Nian} X.-N.~Wang, M.~Gyulassy, Phys.\,Rev.\,Lett.~{\bf 68} (1992) 1480.
\bibitem{Denes} D.~Moln\'{a}r, nucl-th/0503051.
\bibitem{consti} A.~Dainese, C.~ Loizides, G.~Pai\'{c}, Eur.\,Phys.\,J.~{\bf C\,38} (2005) 461.
\bibitem{NCQMolVol} D.~Moln\'{a}r and S.\,A.~Voloshin, Phys.\,Rev.\,Lett.~{\bf 91} (2003) 092301.
\bibitem{piKp} J.~Adams \emph{et\,al.} (STAR Collaboration), Phys.\,Rev.\,Lett.~{\bf 92} (2004) 112301. 
\bibitem{strangeProd} J.~Adams \emph{et\,al.} (STAR Collaboration), Phys.\,Rev.\,Lett.~{\bf 92} (2004) 182301.
\bibitem{omegaFlow} J.~Adams \emph{et\,al.} (STAR Collaboration), Phys.\,Rev.\,Lett.~{\bf 95} (2005) 122301.
\bibitem{phi} J.~Adams \emph{et\,al.} (STAR Collaboration), Phys.\,Lett.~{\bf B\,612} (2005) 181.
\bibitem{Xin} X.~Dong, S.~Esumi, P.~Sorensen, N.~Xu and Z.~Xu, Phys.\,Lett.~{\bf B\,597} (2004) 328.
\bibitem{hydro} P.~Huovinen, private communication (2004).
\end{thebibliography}
\end{document}